\documentclass[aps,twocolumn,floatfix, showkeys, superscriptaddress, nofootinbib]{revtex4-1}
\usepackage{amsmath}
\usepackage{graphicx,bm,hhline}
\usepackage{booktabs}
\DeclareMathAlphabet{\pazocal}{OMS}{zplm}{m}{n}

\newcommand{\bcri}{{\bm R}_i}
\newcommand{\bcrj}{{\bm R}_j}
\newcommand{\bcr}{{\bm R}}

\newcommand{\br}{{\bm r}}
\newcommand{\bx}{{\bm x}}
\newcommand{\bxp}{{\bx}^\prime}

\newcommand{\brp}{{\br}^\prime}

\newcommand{\wig}[1]{\mathrel{\hbox{\hbox to 0pt{\lower.6ex\hbox{$\sim$}\hss}\raise.4ex\hbox{$#1$}}}}

\newcommand{\lp}{l^\prime}

\newcommand{\lpp}{l^{\prime\prime}}

\makeatletter
\renewcommand*\env@matrix[1][\arraystretch]{%
  \edef\arraystretch{#1}%
  \hskip -\arraycolsep
  \let\@ifnextchar\new@ifnextchar
  \array{*\c@MaxMatrixCols c}}
\makeatother

\begin{document}
\title{Real-Space Green's functions for Warm Dense Matter}

\author{M. Laraia}
\affiliation{Los Alamos National Laboratory, P.O. Box 1663, Los Alamos, NM 87545, U.S.A.}
\affiliation{School of Physics and Astronomy, University of Minnnesota, Minneapolis, MN 55455, U.S.A.}

\author{C. Hanson}
\affiliation{Los Alamos National Laboratory, P.O. Box 1663, Los Alamos, NM 87545, U.S.A.}

\author{N. R. Shaffer}
\affiliation{Los Alamos National Laboratory, P.O. Box 1663, Los Alamos, NM 87545, U.S.A.}

\author{D. Saumon}
\affiliation{Los Alamos National Laboratory, P.O. Box 1663, Los Alamos, NM 87545, U.S.A.}

\author{D. P. Kilcrease}
\affiliation{Los Alamos National Laboratory, P.O. Box 1663, Los Alamos, NM 87545, U.S.A.}

\author{C. E. Starrett}
\email{starrett@lanl.gov}
\affiliation{Los Alamos National Laboratory, P.O. Box 1663, Los Alamos, NM 87545, U.S.A.}

\date{\today}
\begin{abstract}
  Accurate modeling of the electronic structure of warm dense matter is a challenging problem whose solution would allow a better understanding of material properties like equation of state, opacity, and conductivity, with resulting applications from astrophysics to fusion energy research.
  Here we explore the real-space Green's function method as a technique for solving the Kohn-Sham density functional theory equations under warm dense matter conditions.
  We find the method to be tractable and accurate throughout the density and temperature range of interest, in contrast to other approaches.
  Good agreement on equation of state is found when comparing to other methods, where they are thought to be accurate.
\end{abstract}
\pacs{ }
\maketitle

\section{Introduction}

The term ``warm dense matter'' refers to states of matter with densities from a fraction of solid to several times solid density, and temperatures of a fraction of an eV to hundreds of eV.
The electronic structure of warm dense matter is strongly density- and temperature-dependent, seen in effects like pressure ionization as well as the relocalization of electrons with increasing temperature (caused by reduced screening of the nucleus by hot electrons).
This state of matter appears in inertial fusion experiments \cite{haan11,lepape18,gomez14}, basic science experiments \cite{glenzer16}, and in astrophysical objects like white dwarf stars \cite{kowalski07, blouin20}, giant planets \cite{helled20,bethkenhagen20}, and the interior of the sun \cite{bailey15,nagayama19}.

The electronic structure of warm dense matter is routinely calculated using Kohn-Sham density functional theory (DFT)\cite{kohn65,mermin65}, most often using a plane-wave expansion of the electron orbitals.
However, this method has certain drawbacks that limit its applicability across the warm dense matter regime.
The first drawback is a limitation to relatively low temperatures due to prohibitive scaling with temperature when the degeneracy parameter $\Theta=k_BT/E_F$ exceeds unity, where $T$ is the temperature and $E_F$ is the Fermi energy \cite{sjostrom14}.
A second drawback is the use of pseudo-potentials to represent the screening of the nucleus by core electrons.
These must be carefully created and results checked for sensitivity with respect to them \cite{witte17,witte18}, since even core electrons are affected by the plasma environment in warm dense matter.

In this work, we consider an alternative approach to the electronic structure of warm dense matter based on a real-space Green's Functions (RSGF) solution of the DFT equations.
The RSGF method sidesteps some of the problems encountered with plane-wave DFT when applied to warm dense matter.
Notably, it is based on the one-body Green's function rather than one-body orbitals, which allows more favorable scaling at elevated temperature.
It also uses a multi-center expansion of the Green's function with spherical harmonics, which circumvents the need for pseudo-potentials and allows genuine all-electron calculations.

RSGF is a method with a long history, and in particular has been very useful for the optical properties of materials at standard temperature and pressure, including solids and molecular systems\cite{ankudinov98,rehr00,zabinsky95,rehr92,abrikosov97,wang95,ebert99}.
The x-ray spectra of warm dense matter using RSGF and related methods have also been explored \cite{mattern12, mattern13, peyrusse08}.

Two of us (CES and NRS) recently explored the use of the closely related Korringa-Kohn-Rostoker Green's Function (KKR-GF) \cite{korringa47, kohn54, ebert11} method for the electronic structure of warm dense matter \cite{starrett20ms}.
Both methods, KKR-GF and RSGF, fall under the broader theoretical concept known as multiple scattering theory \cite{ebert11}.
The difference between KKR-GF and RSGF comes down to whether one approximates the infinite plasma using a periodically repeating system (KKR-GF) or by a finite cluster (RSGF).
While RSGF is appealing for its simplicity and avoidance of artificially imposed periodicity, it is not obvious {\it a priori} how large the cluster must be to attain converged electronic structure.
Further, if converged calculations are possible for warm dense matter, it is not obvious if the required cluster sizes are so large to make the method impractical.


In this contribution we demonstrate that RSGF is in fact a viable and accurate method for predicting the equation of state (EOS) and electronic structure of warm dense matter.
We also discuss the advantages of RSGF over KKR-GF.
In short, these are simplicity in implementation and avoidance of $k$-space integrations, which makes the method better suited for calculations of spectral quantities like densities of state (DOS) and optical constants.
Convergence of the EOS with respect to cluster size is demonstrated, and shown to be more computationally efficient than equivalent KKR-GF calculations.
Comparison is then made with an average atom model on EOS and DOS.
It is found that while RSGF and the average atom models agree on EOS at low densities and high temperatures, the DOS remains qualitatively different due to broadening of states by the plasma environment treated in RSGF but neglected in the average atom.
We develop a perturbative model of this effect to broaden the average atom DOS.
This improves the qualitative agreement between the models, mainly by destroying the average atom's artificially sharp valence states and continuum resonances.

\section{Real-Space Green's Functions}

In multiple scattering theory, space is divided into cells, each containing a special point called its center.
Each of these centers is taken to be the origin of its cell, and the multi-center electronic structure problem is recast as many single-center ones.
Done na\"ively, the boundary conditions on each cell would be very complicated due to both the cell geometry and the need to match wavefunctions in neighboring cells.
By representing the electronic structure in terms of the complex-energy one-body Green's function, $G(\bx, \bxp, z)$, one can instead solve each cell with simple boundary conditions, (e.g., free-particle).
By also calculating the scattering $t$-matrix for each cell, these single-center solutions can be corrected to obtain the full multi-center Green's function by solving a Dyson equation.

In practice, the method uses a double spherical harmonic expansion of the Green's function, one for each position argument.
If $\bx$ and $\bxp$ lie respectively in cells $n$ and $n'$ with centers $\bcr^n$ and $\bcr^{n'}$, then one writes $\bx = \br + \bcr^n$ and $\bxp = \brp + \bcr^{n'}$ and expands the $\br$ and $\brp$ dependence in spherical harmonics about their respective centers.
The full Green's function decomposes into single-site and multi-site terms
\begin{multline}
  G(\br+\bcr^n ,\brp+\bcr^{n'},z)= G^{ss}(\br+\bcr^n ,\brp+\bcr^{n'},z) \\
  + G^{ms}(\br+\bcr^n ,\brp+\bcr^{n'},z)
  \label{gfa}
\end{multline}
The single site Green's function is given by
\begin{multline}
  G^{ss}(\br+\bcr^n ,\brp+\bcr^{n'},z)= \\
  2m_e\delta_{nn'}
  \sum_{L=0}^\infty H^{n,\times}_L(\br_>,z) R^{n}_L(\br_<,z)
  \label{gfss}
\end{multline}
which represents the electronic structure of the system as if each cell were considered separately and then superposed.
The multi-site Green's function 
\begin{multline}
  G^{ms}(\br+\bcr^n ,\brp+\bcr^{n'},z)= \\
  2m_e\sum_{LL'}^\infty R^{n}_L(\br,z)
  {\cal G}^{nn'}_{LL'}(z)   
  R^{n'\times}_{L'}(\brp,z)
  \label{gfms}
\end{multline}
then corrects the electronic structure to account for interference from other sites.
Here, $R^n_L(\br,z)$ and $H^n_L(\br,z)$ are the complex-energy solutions of the Kohn-Sham Schr\"odinger or Dirac equations in cell $n$, separated in spherical coordinates about $\bcr^n$.
Non-relativistically, these are Pauli spinors with angular momentum numbers $L=(l,m)$; relativistically, these are Dirac spinors with angular momentum numbers $L=(\kappa,m)$.
The superscript $\times$ means to take the complex conjugate of the angular part of the non-relativistic solution or to take the Hermitian conjugate of the spin-angular part of the relativistic solution.
In the following, we only show results for non-relativistic calculations.
The notation $\br_>$ ($\br_<$) means to take $\br$ or $\br'$ according to which one is greater (lesser) in magnitude.
Throughout, we use Hartree atomic units with $\hbar = 4\pi\epsilon_0 = e^2 = 1$, leaving $m_e$ symbolic for easy conversion to Rydberg units.
For other normalization and sign conventions, see Reference~\cite{starrett20ms}.

The key quantities of multiple scattering theory are the ${\cal G}^{nn'}_{LL'}$ in Eq.~\eqref{gfms}, called the structural Green's function matrix elements.
These make up the structural Green's function matrix $\bm{\mathcal G}$, which is calculated from the structure constants matrix $\bm{\mathcal G}_0 $ and the $\bm t$-matrix using Dyson's equation
\begin{equation}
  {\bm{\mathcal G}}(z)=
  {\bm{\mathcal G}}_0(z)
  \left[I-
    {\bm{t}}(z)
    {\bm{\mathcal G}}_0(z)
  \right]^{-1}
  \label{sgf}
\end{equation}
This expression is known as the fundamental equation of multiple scattering theory and is solved by matrix inversion.
The size of the matrix is $N(l_{max}+1)^2$, where $N$ is the number of cells, and $l_{max}$ is the maximum value of the orbital angular momentum quantum number that occurs in Eq.~\eqref{gfms}.
While in principle $l_{max}$ goes to infinity, in practice it is accurate to use small values like 2 or 3 \cite{starrett20ms}.

The structure constant matrix elements $G_{0,LL^\prime}^{nn^\prime}$ depend on the positions of the expansion centers $\{\bcr^n\}$ and the energy $z$.
For periodic systems, one uses the Ewald technique and a Fourier space integration and the result is the so-called Korringa-Kohn-Rostoker Green's function (KKR-GF) method \cite{korringa47,kohn54}.
For a finite cluster of centers one can directly evaluate the structure constants in real space \cite{zabloudil_book}
\begin{multline}
  G_{0,LL'}^{nn'}(z) = - 4 \pi i p \sum_{L''}
  i^{l-\lp-\lpp} h_{\lpp}^+(p R_{nn^\prime}) \\
  \times
  C_{LL''}^{L'} Y_{L''}(\hat{\bm R}_{nn'})
  \label{rss}
\end{multline}
which is the Real-Space Green's function (RSGF) method.
Here, $Y_{L}$ is a spherical harmonic including the Condon-Shortley phase, ${\bm R}_{nn'}= {\bm R}_n - {\bcr}_{n'}$ is the vector connecting centers $n$ and $n'$, $p=\sqrt{2mz}$ is the complex momentum lying in the upper-right quadrant of the $z$ plane, $h^+_l$ is the spherical Hankel function, and 
\begin{equation}
  C_{L L"}^{L'}  \equiv
  \int d\hat{\bm r} 
  Y_{L}(\hat{{\bm r}})
  Y_{L'}^*(\hat{{\bm r}})
  Y_{L''}(\hat{{\bm r}})
\end{equation}
are Gaunt coefficients, defined here to be real-valued.
As there is no Ewald sum or Fourier space integration, the RSGF expression, Eq.~\eqref{rss}, is much simpler to implement than the corresponding KKR expression (c.f. the Appendix of Ref.~\cite{starrett20ms}). 

The RSGF method is strictly valid only for finite clusters of centers, whereas the KKR-GF method is valid for any periodic system.
For the electronic structure of bulk disordered systems like plasmas, either method is approximate and introduces finite size effects depending respectively on the number of particles in the cluster or in each periodic image.
Intuitively, it would seem the finite-size effects of the KKR-GF method might be less severe in general.
However, if one takes the cluster to be sufficiently large and focuses on the electronic structure of cells near the center of the cluster, then the RSGF $\bm{\mathcal G}$ can be considered to be an approximation to that of KKR-GF.
The essence of the approximation is that long-range structure far from the center of the cluster does not significantly influence the electronic structure there.
This cluster approximation should be especially well-suited to disordered plasmas, as opposed to ordered crystals, because coherent Bloch states that are possible in crystals are destroyed by the ionic disorder.

\begin{figure}
  \begin{center}
    \includegraphics[scale=0.6]{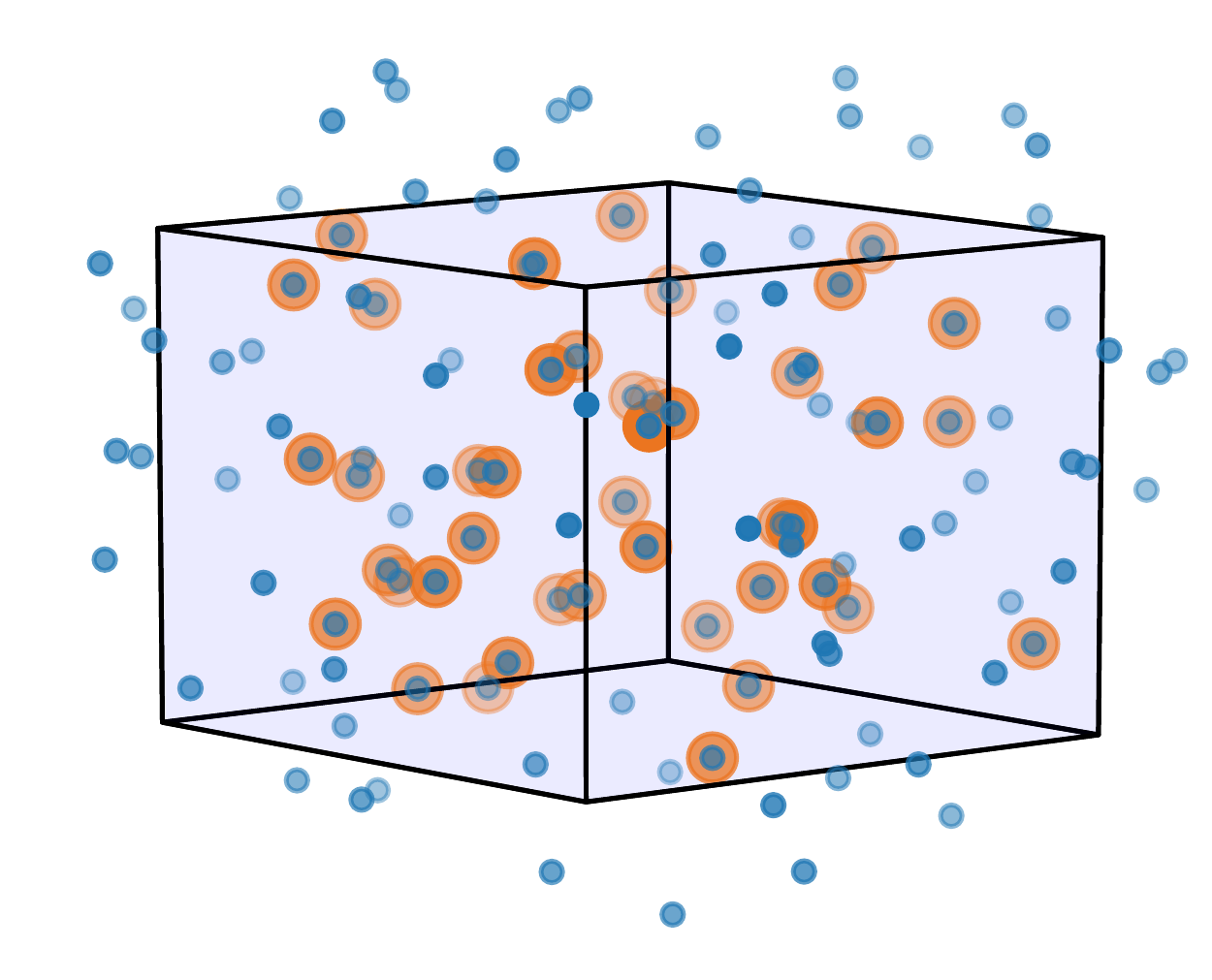}
  \end{center}
  \caption{(Color online) Illustration of centers included in the real-space Green's function cluster approximation.
    The computational cube is shown as the shaded box, centers are shown as blue (dark) circles, and those inside the box are surrounded by orange (light) shaded spheres.
  }
  \label{fig_clu}
\end{figure}

\section{Results}

\subsection{Physical Model, Convergence, and Timing}

We adopt the same physical model as presented in reference \cite{starrett20ms}.
We consider a periodic computational cube with $N$ nuclei of charge $Z$ and $NZ$ electrons.
The positions of the nuclei are provided by the pseudoatom molecular dynamics (MD) model \cite{starrett15a}.
In the results shown here, we use non-relativistic DFT in the temperature-dependent local density approximation (LDA) \cite{ksdt}.
The tessellation of space is carried out using the power tessellation technique \cite{alam09}, and centers not corresponding to nuclear positions are added to minimize the volume not in the muffin-tin spheres, as in \cite{starrett20ms}.
\begin{figure*}
  \begin{center}
    \includegraphics[scale=0.6]{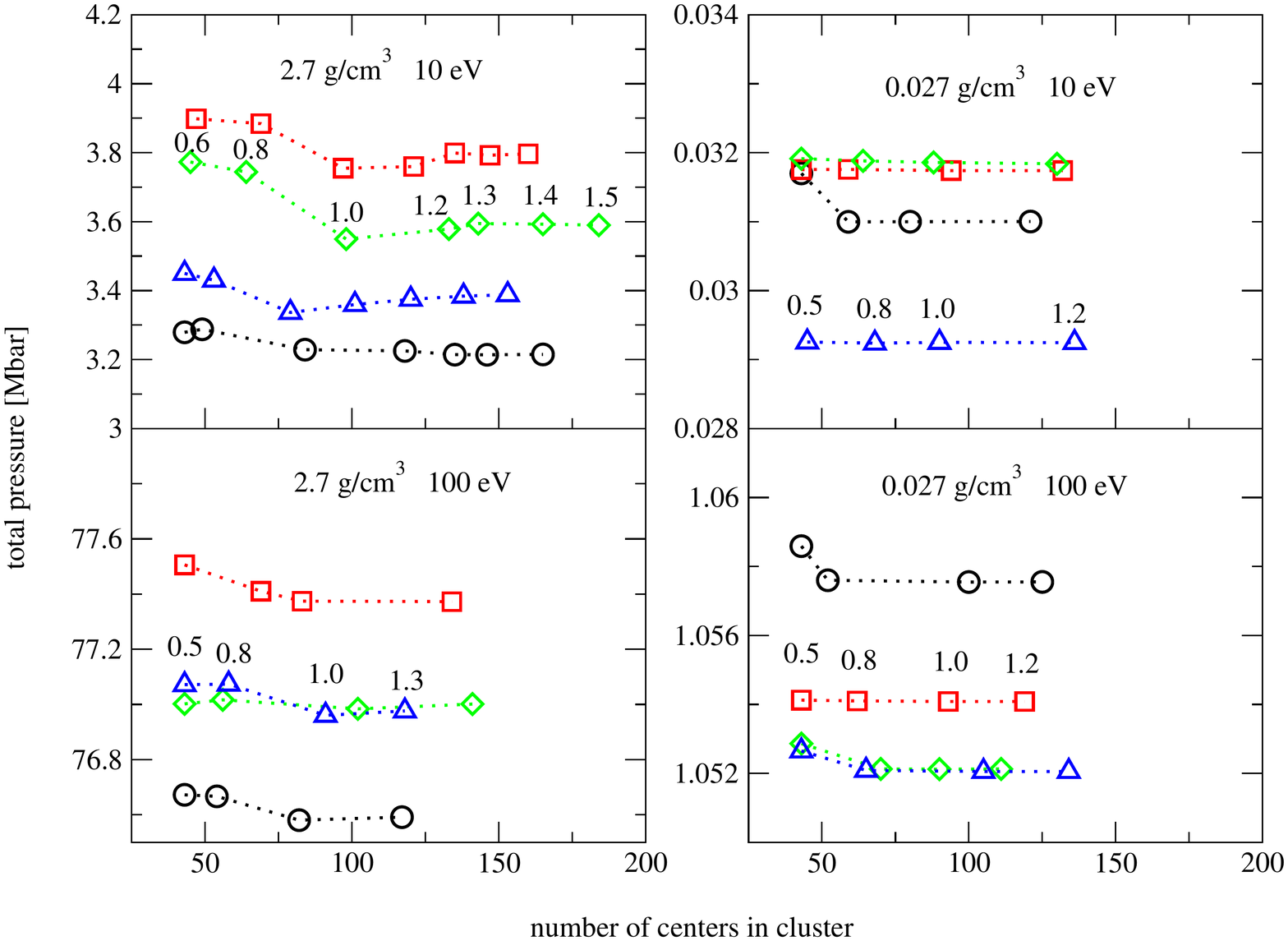}
  \end{center}
  \caption{(Color online) Convergence of the total pressure of aluminum plasmas with respect to the number of particles in the correlation sphere.
    For each temperature and density condition, four distinct lines are shown corresponding to four MD frames.
    On one line the size of the correlation radii in units of the ion-sphere radius are given.
    The computational cube contains 8 nuclei and 35 extra centers, therefore the minimum number of centers in this cluster approximation is 43.}
  \label{fig_np}
\end{figure*}

The difference to the model presented in reference \cite{starrett20ms} is that the structural Green's function matrix is calculated using the RSGF cluster approximation.
To define the cluster, we include at a minimum all centers in the computational cube.
We further include any centers outside that cube but within a fixed distance from any center in the computational cube, illustrated in Figure~\ref{fig_clu}.
This fixed distance we call the \emph{correlation radius}.
The larger the correlation radius, the more centers enter into the calculation of the structural Green's function matrix, Eq.~\eqref{sgf}, and the better converged the multi-site contribution to the electronic structure.
The correlation radius is one of the main parameters to check convergence with in RSGF, taking the place of $k$-point sampling density in KKR-GF.

In Figure~\ref{fig_np}, we show numerical convergence of the EOS with respect to the correlation radius for a range of aluminum plasmas.
It is seen that the size of the correlation radius necessary for convergence of the pressure depends on the density and temperature.
At higher temperatures, the multiple scattering correction is less important, as pointed out in Ref.~\cite{starrett20ms}.
This is because the electrons of high-temperature plasmas have higher kinetic energies on average and therefore it is reasonably accurate to assume free-electron boundary conditions on each site, equivalent to ignoring $G^{ms}$ in Eq.~\eqref{gfa}.
At low densities we also find the multi-center Green's function to be less important because free-electron boundary conditions become exact in the limit of isolated atoms.

\begin{table}[]
  \begin{center}
    \bgroup
    \def\arraystretch{1.5}%
    \begin{ruledtabular}
      \begin{tabular}{ccc cc}
        T [eV] & $\rho$ [g/cm$^3$] & RSGF & $r_c$ & No MS \\
        \hline
        10      & 2.7               & 15.7 & 1.3   & 2.7   \\
        10      & 0.027             & 11.4 & 0.8   & 2.9   \\[0.2cm]
        100     & 2.7               & 14.4 & 1.0   & 3.3   \\
        100     & 0.027             & 9.5  & 0.8   & 2.6   \\
      \end{tabular}
    \end{ruledtabular}
    \egroup
  \end{center}
  \caption{Wall time in minutes to convergence for one MD frame, run on an 18-core CPU.
    Times shown for RSGF solution with $l_{max}=2$ and with no multiple scattering solution (No MS).
    The size of the correlation sphere radius ($r_c$) is also given, in units of the ion-sphere radius.  }
  \label{tab_t}  
\end{table}
Table~\ref{tab_t} shows timing results for the same conditions shown in Figure~\ref{fig_np}.
For each density, the correlation sphere radius was taken to be the smallest converged value.
In all cases, the time to obtain an SCF solution is modest.
Comparing the solid density timings to the KKR-GF timings reported in Ref.~\cite{starrett20ms}, we see approximately a fourfold speed up in the RSGF method due its more economical calculation of the structure constants.
This is helped by the fact the correlation radius in RSGF is a fine-grained parameter to determine convergence, compared to refining the $k$-space integration mesh in KKR-GF.
That is, one can gradually increase $r_c$ in RSGF in order to obtain converged results without wasted effort, whereas going from a $2^3$-point mesh to a $3^3$-point one in KKR-GF represents a big jump in expense.

\subsection{Equation of State}

\begin{table*}[]
  \begin{center}
    \bgroup
    \def\arraystretch{1.5}%
    \begin{ruledtabular}
      \begin{tabular}{cc ccccccc}
        T [eV] & $\rho$ [g/cm$^3$] & RSGF  & KKR-GF      &  No MS & difference & {\texttt Tartarus}  & difference \\
        \hline
        10 & 2.7      & 3.35 (0.20) & 3.36 (0.18) &  3.64 (0.25)        & 8.7\%    & 2.93 & -13\%  \\
        10 & 0.27  &  0.291 (2.3$\times$10$^{-2}$) &  & 0.300 (2.3$\times$10$^{-2}$) & 3.1\%     & 0.271  & -6.8\% \\
        10 & 0.027 & 3.12$\times$10$^{-2}$ (5.7$\times$10$^{-4}$) && 3.17$\times$10$^{-2}$ (3.3$\times$10$^{-4}$)   & 1.6\%  & 3.11$\times$10$^{-2}$  & -0.3\% \\[0.2cm]
        100  &   2.7   & 76.7 (0.51) & 76.6 (0.35)   &  77.0 (0.53)   &  0.3\%  &  75.4 & -1.7\% \\
        100  &   0.27 &  9.11 (2.2$\times$10$^{-2}$)   &&  9.12 (2.3$\times$10$^{-2}$) & 0.1\% & 9.03  & -0.9\% \\
        100  &  0.027 & 1.05 (7.5$\times$10$^{-3}$) &&  1.05 (3.4$\times$10$^{-3}$)  & 0\%  & 1.05  & 0\%\\
      \end{tabular}
    \end{ruledtabular}
    \egroup
  \end{center}
  \caption{Total pressures in Mbar for aluminum plasmas from the RSGF method, compared to KKR-GF results of reference \cite{starrett20ms}, and to the {\texttt Tartarus} average atom model \cite{starrett19}, as well as to a calculation with the multiple scattering Green's function set to zero (No MS).
    Also shown, beside the No MS and {\texttt Tartarus} columns are the percentage differences of that model's pressure compared to the RSGF result.
    The numbers in brackets are one standard deviation.}
  \label{tab_p}  
\end{table*}

\begin{figure*}
  \begin{center}
    \includegraphics[trim=0 6.5cm 0 0, clip, width=0.8\textwidth]{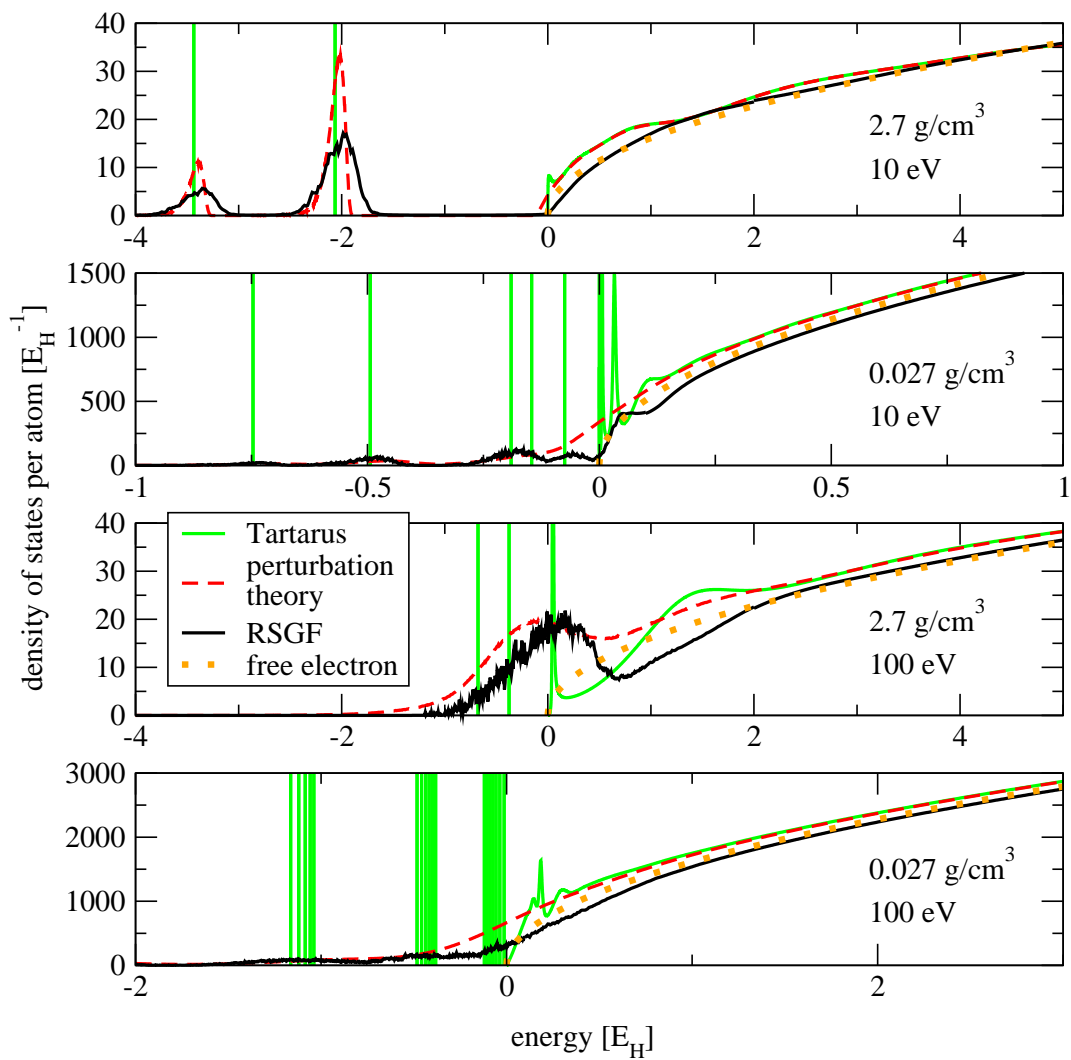}
  \end{center}
  \caption{(Color online) Density of states for aluminum plasmas.
    Shown are the results from the RSGF method, where the ``noise'' is due to averaging over MD frames.
    Also shown is the DOS from the {\texttt Tartarus} average atom model, and the DOS resulting from using a microfield and perturbation theory to broaden the {\texttt Tartarus} result.
    For the {\texttt Tartarus} result, the vertical lines for negative energy values correspond to the discrete core states.}
  \label{fig_dos}
\end{figure*}
Table \ref{tab_p} presents total pressures from the RSGF method for aluminum plasmas.
The pressure is evaluated as a time average over MD configurations with 8 nuclei and 35 extra centers \footnote{See Ref.~\cite{starrett20ms} for a convergence study on the number of extra centers.} in the computational volume.
First, we compare to the KKR-GF results of Ref.~\cite{starrett20ms} and find good agreement, validating that the RSGF finite-cluster approximation is a good one.
Note that we have used correlation sphere radii sufficiently large that larger values do not change the results significantly (1.3 at 2.7 g/cm$^3$ and 0.8 at 0.027 g/cm$^3$, each units of the ion-sphere radius for that density).
The column labeled ``No MS'' in the table refers to a calculation in which we have set the multiple scattering Green's function to be zero.
Comparing this calculation to the RSGF results reveals excellent agreement at 100 eV, for all densities, affirming that multiple scattering effects are small at high temperature.
At 10 eV, differences increase with density as expected due to the increasing overlap of valence electrons of nearby atoms.

In the table we also compare to the {\texttt Tartarus} average atom model.
The link between the multiple scattering model and the average atom was explored in Ref.~\cite{starrett18}.
In short, the average atom model is obtained from RSGF by ignoring the multi-site Green's function, setting the cells to be identical spheres with the ion-sphere radius, and solving the Poisson equation inside each sphere independently.
The average atom approximation is more severe than just neglecting multiple scattering.
Indeed, the agreement of total pressure between {\texttt Tartarus} and RSGF is generally worse than the agreement between RSGF and ``No MS'', but it is still quite reasonable at higher temperatures or lower densities.
The fair agreement at high density and temperature might be surprising, since the MD simulation allows nuclei to get closer together than the radius of the ion-sphere.
On the one hand, the average atom approximation ought to break down due its neglect of electron overlap between atoms.
On the other hand, the EOS at high temperature is controlled mainly by high-energy, nearly free electrons, whose contribution to the EOS is essentially correct even in the average atom model.
The same cannot be said for electrons in the lower energy valence states, which are poorly treated by the average atom model.
Quantities that interrogate the valence, such as the density of states or absorption spectra, will display strong disagreement between average atom and RSGF calculations, even at conditions where the two approaches agree in the EOS.

\subsection{Density of States}

In addition to the simplicity and rapid convergence of RSGF, another advantage it holds over the KKR-GF method comes in computing spectral quantities like the density of states (DOS).
One is generally interested in the DOS at real energies, for which one needs to evaluate the Green's function on or very near the real energy axis, where it is a rapidly varying function of energy.
In the KKR-GF method, this requires many $k$-space integration points to accurately calculate the structural Green's function (more than needed for EOS), which is very costly.
In contrast, we find the RSGF cluster sizes needed for converged DOS calculations to be no greater than what is needed for EOS in practice.
This makes RSGF well-suited to the calculation not only of DOS, but for optical constants as well, which are intrinsically more expensive to evaluate \cite{rehr00}.

In Figure \ref{fig_dos}, the density of states in the valence region is shown for aluminum plasmas.
Looking first at the RSGF results at 10 eV and solid density, we see two distinct bands at negative energy, corresponding to the 2$s$ and 2$p$ bound states, as well as a free-electron-like continuum for positive energies.
For lower densities or higher temperatures, the valence states broaden considerably to the point of merging with the continuum.

It is useful to compare the DOS with that from the {\texttt Tartarus} average atom model.
In average atom models, the bound orbitals are at specific energies and are $\delta$-distributed in the DOS, no matter how weakly bound.
The eigenvalues of the average atom bound spectrum do tend to correspond to bands in the RSGF calculation, but because of the lack of an explicit plasma environment, there is no broadening of these states.
The case of 0.027 g/cm$^3$ and 100 eV highlights this defect of the average atom model, showing series of atomic-like bound states and multiple continuum resonances, all of which are completely washed out in the RSGF calculation.

The average atom model is missing physics that would lead to the broadening of these bound state features seen in RSGF.
The missing effect is the electric potential fluctuations from atom to atom as well as in time due to the plasma environment.
If one knew the statistical distribution of electric potential felt by a bound electron, one could use perturbation theory to approximate the distribution of eigenvalue shifts due to the fluctuating potential, which would appear as a broadening of the bound state.

The distribution of electric potentials can be approximated using the same pseudoatom MD model used to generate the RSGF configurations \cite{starrett15a}.
In that model, the total potential is given by a superposition of spherically symmetric pseudoatom potentials $V^{PA}(r)$, 
\begin{equation}
  V(\bx) = \sum_j V^{PA}(|\bx-\bcrj|)
\end{equation}
As a model for the plasma's perturbing effect on the average atom model, we consider the deviation of the potential near atom $i$ from the average atom model potential $V^{AA}(r)$
\begin{align}
  \Delta V_i(\br)
  &= V(\br+\bcri) - V^{AA}(r) \\
  &= V^{PA}(r) - V^{AA}(r) + \sum_{j\ne i} V^{PA}(|\br + \bcri - \bcrj|)
\end{align}
where $\br$ is the position vector measured from the atom's location.
For $r < |\bcri-\bcrj|$, the first term in a multipole expansion of this potential near the atom is a constant
\begin{align}
  \Delta V_i(\br) & \approx \Delta V_i(0) \nonumber \\
                         &=V^{PA}(0)-V^{AA}(0)+ \sum_{j \ne i} V^{PA}(|\bcri-\bcrj|)
\end{align}
We take this constant as an approximation to the plasma potential experienced by the electrons within atom $i$ in the configuration $\{\bcrj\}$.
A statistical distribution of plasma potentials is accumulated over each atom and time step of an MD simulation.  We further assume that the potential distribution is centered about the average atom potential.
Treating $\Delta V$ as a random perturbation to the average atom Hamiltonian leads, at first order, to a shift of the average atom eigenvalues by the constant $\Delta V$.
The distribution of these shifts leads to a broadening of the average atom DOS, calculated by convolving the normalized $\Delta V$ distribution with the average atom DOS.

In Fig.~\ref{fig_dos}, the red dashed lines show the result of applying the monopole broadening perturbation theory to the average atom DOS.
The qualitative agreement with RSGF is much improved, with $\delta$-distributed bound states and sharp continuum resonances broadened into bands of states.
The model is not quantitative, but still gives reasonably realistic valence structure in spite of the crude approximations involved.

The monopole approximation could be improved either by going to higher orders in the expansion of $\Delta V(\br)$ or by expanding about a point other than the origin, for instance, each orbital's mean radius.
Either approach would predict more broadening, and this broadening would be orbital-dependent rather than constant.
For relatively deep valence states, like the 2$s$ and 2$p$ features at 2.7 g/cm$^3$ and 10 eV, this would likely improve agreement with RSGF. However, for the other conditions shown in Fig.~\ref{fig_dos}, where the valence structure is closer to the threshold, the accuracy of the broadening model is probably limited by perturbation theory rather than the details of the plasma potential approximation.
This is because perturbation theory is likely not convergent for near-threshold valence states, where the calculated bands are wider than the spacing between eigenstates.
A direct re-diagonalization of the average atom Hamiltonian is possible in principle, but to do self-consistently is not very practical, especially if considering higher multipole effects that break the spherical symmetry of the average atom.

\section{Conclusions and discussion}

In this work, the real-space Green's function (RSGF) method has been demonstrated to be an effective technique for solving the Kohn-Sham DFT equations for warm dense matter.
The method is tractable for plasmas at any temperature, since it avoids the calculation of explicit eigenstates that makes orbital-based DFT methods expensive at high temperature.
RSGF relies on a cluster approximation, where it is assumed that the positions of the nuclei far from the center of the cluster do not affect the electronic structure near the center.
A numerical study of the convergence of the pressure with respect to the cluster size affirms the accuracy of this approximation, even with modest cluster sizes.
Comparisons of the RSGF model's predicted equation of state showed excellent agreement with KKR-GF results at high densities, while at low densities we find agreement with the average atom model {\texttt Tartarus}.
An examination of the density of states and comparison with the {\texttt Tartarus} model highlighted important broadening physics lacking from the average atom model but accounted for in RSGF, even at conditions where the equation of state is in good agreement.
This led us to propose a simple model for the broadening of average atom bound states and resonances based on the qualitative effect of plasma disorder lacking in average atom models.
We conclude that RSGF is an accurate and practical method for solving the Kohn-Sham DFT equations in warm dense matter conditions and is also useful for highlighting shortcomings and possible refinements to the average atom model.

\section*{Acknowledgments}
This work was performed under the auspices of the United States Department of Energy under contract DE-AC52-06NA25396.

\appendix

\bibliographystyle{unsrt}
\bibliography{phys_bib}

\end{document}